# Screening for electrically conductive defects in thin functional films using electrochemiluminescence

Harley Quinn,[1] Wenlu Wang,[1] Jörg G. Werner,[1,2] Keith A. Brown[1,2,3,*]

[1]Division of Materials Science & Engineering, Boston University, Boston, MA, 02215, USA

[2]Department of Mechanical Engineering, Boston University, Boston, MA, 02215, USA

[3]Physics Department, Boston University, Boston, MA, 02215, USA

[*]Correspondence to be addressed to *brownka@bu.edu*

**Abstract**

Multifunctional thin films in energy-related devices often must be electrically insulating where a single nanoscale defect can result in complete device-scale failure. Locating and characterizing such defects presents a fundamental problem where high-resolution imaging methods are needed to find defects, but imaging with high spatial resolution limits the field of view and thus the measurement throughput. Here, we present a novel high-throughput method for detecting sub-micron defects in insulating thin films by leveraging the electrochemiluminescence (ECL) of luminol. Through a systematic study of reagent concentrations, buffers, voltage, and excitation time, we identify optimized conditions at which it is possible to detect features with areas ~500 times smaller than the area interrogated by a single pixel of the camera, showing high-throughput detection of sub-micron defects. In particular, we estimate the minimum detectable features to be lines as narrow as 2.5 nm in width and pinholes as small as 35 nm in radius. We further explore this method by using it to characterize a nominally insulating phenol film and find conductive defects that are cross-correlated with high-resolution atomic force microscopy to provide feedback to synthesis. Given the inherent parallelizability and scalability of this assay, it is expected to have a major impact on the automated discovery of multifunctional films.

**Introduction**

Thin multifunctional films are ubiquitous in energy-related devices including fuel cells,[1] batteries,[2–6] and photovoltaics.[7,8] A key requirement for such films is that they allow the transport of different species to be independently managed. For example, solid-state battery electrolyte films require that ions can readily pass through the film while electrons cannot. The requirement for electrical insulation is particularly insidious in the context of scalability because a single nanoscale defect in a device that has square centimeters of functional area can critically impact device performance.[3,4] Detecting the presence of a defect, for example, through leakage current or short circuits, is straightforward, but identifying its location is essential to determining its origin and finding synthesis and processing conditions that mitigate such defects. However, locating and characterizing defects presents a fundamental problem where high-resolution imaging methods are needed to find defects, but imaging with high spatial resolution typically comes with the tradeoff of limiting the field of view. Current techniques used for defect detection include direct inspection techniques such as atomic force microscopy (AFM)[9–11] and scanning electron microscopy (SEM).[11–13] In addition to these methods to identify defects through their structure, there are also methods for directly measuring defects through their functional signature such as scanning electrochemical microscopy (SECM), which is a contactless and high-resolution method for measuring local electrochemical activity.[14–16] Early work showed that this method can be used to quantify film conductivity,[14] and then this was extended to a mapping approach to measure substrate conductivity across 800 μm wide regions.[15] Two drawbacks of probe-based approaches, however, are that data collection is a serial process and resolution is constrained by the ultramicroelectrode tip radius together with the tip-sample separation,[16] making it challenging and prohibitively time consuming to map large regions at a fine resolution.

In contrast with methods for serially mapping the properties of functional films with spatial resolution commensurate with the defects of interest, when considering the identification and mapping of sparse microscopic defects on macroscopic samples, it would be preferrable to amplify defects so that they can be rapidly identified with low-resolution tools. For example, the reductive growth of silver can amplify the optical signature of small defects,[17] but it would be preferrable to employ a method that is non-destructive so that further analysis could be performed once defects are located. Considering these requirements, optical methods that feature a dark-field readout where defects generate light have the advantages of being inherently parallelizable, compatible with low magnification imaging for large-area screening, and the lack of signal generated on correctly functioning films vastly facilitates signal analysis. Furthermore, conventional limitations of optical imaging such as the diffraction limit do not present a challenge when the goal is to make defects appear large. When considering a process for generating light using electrically conductive defects, the electrochemiluminescence (ECL) of luminol stands out as a widely used method.[18–21] In addition to its widespread use in forensics and biology,[18–21] it has also been used as a way to map ECL on surfaces,[22] quantify variations in catalytic activity,[23] and even facilitate the readout of human fingerprints.[24] However, the use of luminol-based ECL to explore conductive defects on nominally insulating films has not been shown and there are open questions as to the role of processing conditions and what resolving power and throughput can be achieved using this method.

Here, we show that luminol ECL can be used to map nanoscale conductive defects on nominally insulating films. Specifically, we explore the hypothesis that applying a potential across a sandwiched electrochemical cell will result in the localized emission of light on exposed regions of the anode (Figure 1A). To turn this hypothesis into an assay, we first perform a systematic study of the ECL conditions including reagent concentrations, properties of the buffer, excitation voltage, and excitation duration. Interestingly, these factors have a subtle interplay with a decrease in ECL occurring at excessive voltages, durations, or luminol concentrations. This optimization in hand, we perform a study of the ECL signal originating from nanoscale dot and line defects as a function of their size and use this information to estimate the minimum resolvable features, which we estimate to be 2.5 nm wide lines or 35 nm radius dots. Finally, we employ this assay to map a number of conductive defects on an electrodeposited ultrathin poly(phenylene oxide) film and show that these can be colocalized with AFM as a higher resolution technique. Collectively, these experiments show that luminol ECL provides a rapid and parallelizable way of observing defects and providing feedback for the development of advanced multifunctional films.

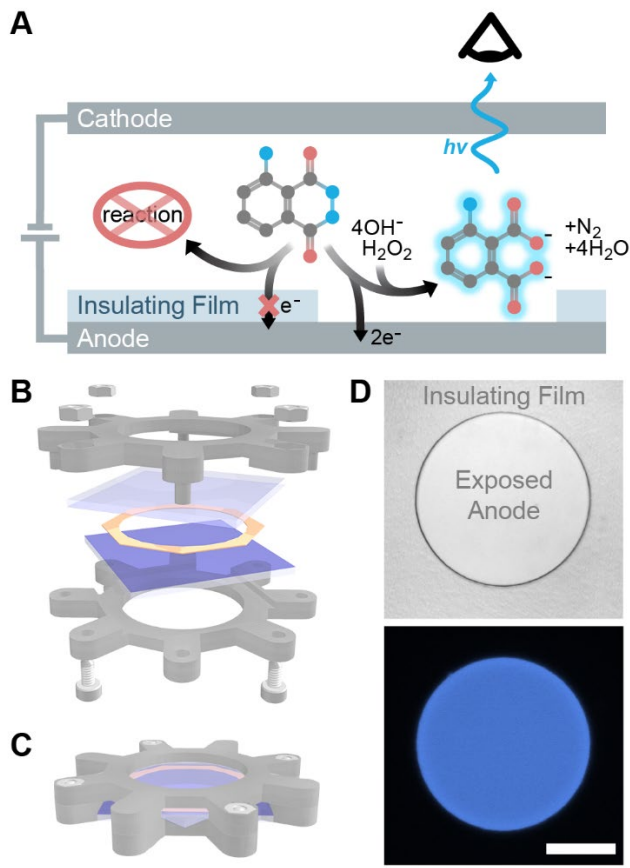

**Figure 1.** (A) Schematic of the electrochemical cell (side view) showing the mechanism of luminol electrochemiluminescence (ECL). (B) Exploded view of electrochemical cell and holder (grey) with screws and nuts (light grey), glass slides (light blue) with ITO coating (medium blue), and spacer (gold). (C) Assembled electrochemical cell. (D) Optical micrographs through the electrochemical cell showing a region of the anode in which a 500 µm diameter circular region is exposed while the rest of the anode is protected by an insulating film. Here, the electrochemical cell is filled with 1 mM luminol, 5 mM $H_2O_2$ pH 12 NaOH solution. The top image shows the region when illuminated externally while the bottom image shows the same region without external illumination but with the exposed anode luminescing through ECL upon the application of a 1.6 V potential across the cell for 2 s. The bottom image was taken in greyscale and false-colored to represent the color seen by eye. Scale bar is 200 µm.

**Experimental**

*Reagents*

All reagents were used as purchased without further purification. For the buffer screening experiment, four buffers were prepared (1) 18 mM sodium borate (ACS grade, ≥99.5%, Sigma-Aldrich) adjusted to pH 10, (2) 41 mM sodium bicarbonate (ACS grade, Fisher Chemical) adjusted to pH 10, (3) 35 mM sodium bicarbonate adjusted to pH 11, and (4) 33 mM disodium phosphate (ACS grade, ≥99.0%, Alfa Aesar) adjusted to pH 12. All buffers were pH adjusted by adding quantities of 100 mM sodium hydroxide (BioXtra, ≥98.0%, Sigma-Aldrich) solution. Luminol sodium salt (≥98%, Sigma-Aldrich) was dissolved in the relevant buffer for each experiment to reach a concentration of 100 mM luminol to make stock solutions which were stored at 5 °C. Working solutions were prepared the same day of experiments by diluting the stock solution with the same buffers and adding hydrogen peroxide immediately prior to use. After buffer screening experiments were completed, a pH 11 sodium bicarbonate buffer was used to prepare a stock solution which was then aliquoted and stored at -18 °C for all subsequent experiments. Working solutions were prepared the same day of experiments by thawing the stock solution, diluting with buffer, and adding hydrogen peroxide immediately prior to use.

Polymethyl methacrylate (PMMA) electron-beam resist (1000 HARP eB 0.3, KemLab), hydrogen peroxide solution 30% w/w in $H_2O$ (29.0 - 32.0%, Sigma), Microposit MF-319 developer (Rohm and Hass), Microposit S1813 G2 positive photoresist (electronic grade, Rohm and Hass), sodium tetraborate decahydrate (ACS grade, ≥99.5%, Sigma), acetone (semi grade, VWR Chemicals), 2-propanol (semi grade, VWR Chemicals), acetonitrile (anhydrous, Fisher Scientific), tetramethylammonium hydroxide pentahydrate (TMAH, ≥97%, Sigma-Aldrich), tetrabutylammonium perchlorate (TBAP, ≥97.5%, Fisher Scientific), silver perchlorate (anhydrous, Fisher Scientific), and diethyl ether (anhydrous, Sigma-Aldrich) were used as received. Developer for the HARP PMMA e-beam resist was prepared by mixing methyl isobutyl ketone (MiBK, microelectronic grade, J.T.Baker) and 2-propanol (IPA, microelectronic grade, J.T.Baker) in a 1:3 ratio v/v.

*Preparation of substrates*

For the microscale experiments, indium tin oxide (ITO)-coated glass slides (25 × 25 × 1.1 mm$^3$, surface resistivity 70 – 100 Ω/sq, part number 703176 – Sigma Aldrich) were coated with Microposit S1813 G2 positive photoresist using a Headway Research spinner (PWM32) at 4,000 rpm resulting in a resist thickness between 1 and 1.5 µm. A series of circles (diameters between 100 and 1500 µm) was patterned onto each slide using a Karl Suss MA6 mask aligner to define multiple separated regions of exposed ITO on one substrate.

For the nanoscale experiments, ITO-coated glass slides were coated with 1000 HARP PMMA e-beam resist (eB 0.3, KemLab) using a Headway Research spinner (PWM32) at 4,000 rpm resulting in a resist thickness between 0.2 and 0.3 µm. A series of lines (between 0.1 and 1 µm wide) or circles (diameters between 0.1 and 1 µm) were patterned onto the slide using a Zeiss Supra 40VP field emission scanning electron microscope (FE-SEM) with a Nanometer Pattern Generation System (NPGS, JC Nabity Lithography Systems) to define multiple separated regions of exposed ITO on one substrate. Lines were designed to be 50 µm long and spaced by 75 µm in the narrow dimension and 50 µm in the long dimension. Circles were positioned in columns that were 75 µm apart and rows that were 100 µm apart. A grid of samples was prepared with exposure doses between 150 and 3000 µC/cm$^2$, but analysis of ECL focused on a single dose and all studied features were interrogated in a single ECL experiment using 5× magnification. Verification of the feature sizes was carried out using the same FE-SEM with an in-lens detector after ECL experiments were performed. A moderate accelerating voltage of 4.40 kV was chosen to mitigate sample charging.

For the phenol films, electrodeposition of an ultrathin layer of dielectric poly(phenylene oxide) onto ITO-coated glass slides was performed in 10 mM acetonitrile. A stoichiometric ratio of TMAH was used to form the oxidizable phenolate with 100 mM TBAP as the supporting electrolyte. For the deposition, a three-electrode system was used with ITO as the working electrode, a platinum wire as the counter electrode, and $Ag/Ag^+$ as the reference electrode. The reference electrode was composed of a silver wire in 0.05 M silver perchlorate and 0.1 M TBAP in acetonitrile and separated from the monomer solution by a Gamry glass frit. Chronoamperometry (CA) was conducted on the solution at 0.1 V vs. $Ag/Ag^+$ for 20 minutes using a Gamry Reference 600+ potentiostat. After deposition, the film was cleaned with pure acetonitrile followed by diethyl ether.

*ECL Imaging*

A working solution composed of 3.2 mM luminol, 5.6 mM $H_2O_2$, in a sodium bicarbonate buffer (35 mM, pH 11) was used in the experiments shown in Figures 3, 4, and 5. For the experiments shown in Figure 2, different luminol concentrations, peroxide concentrations, and buffer compositions were explored. A fluid cell was constructed by placing a laser-cut polyimide spacer (part 2271K72 – McMaster) with a thickness of 177 ± 1 μm onto an ITO-coated slide with a polymer coating. Then, the ITO-coated slide was placed into a 3D-printed mounting (printed using a FormLabs Form2 out of grey resin). Next, 10 μL of the working solution was pipetted onto the center of the slide, a second ITO-coated slide that had no polymer coating was placed into a second 3D-printed mounting and this assembly was combined with the first mounted slide to form a fluid cell.[25] The fluid cell is shown in an exploded-view in Figure 1B and assembled in Figure 1C. After assembly, this fluid cell was secured under an Olympus BX43 optical microscope with a Hamamatsu ORCA-R2 digital camera C10600 and a dark-field filter cube. This provided a direct view of the patterned electrode surface through the fluid cell (Figure 1D).

An Arduino Uno was used to define the timing of the ECL experiments. In particular, it sourced a timed DC voltage to initiate image capture and then subsequently sourced a programmed DC signal that was routed through an analog filter and scaling amplifier (SIM965, and SIM983 – Stanford Research Systems) to the fluid cell such that a negative voltage was applied to the patterned ITO-coated slide while the unpatterned ITO-coated slide was grounded. Images were taken using 5× magnification with camera exposure times that were chosen to be longer than the duration of the voltage applied to the fluid cell. Reference images were taken immediately preceding each measurement with the same camera exposure time. Analysis was conducted using ImageJ (National Institutes of Health). For display purposes, the brightness and contrast of all images were adjusted using the auto brightness contrast function with fine tune adjustments to the brightness and contrast done manually in ImageJ.

**Results and Discussion**

In order to determine whether ECL of luminol could be used to identify the location of electrically conductive regions on substrates, we performed an experiment in which an ITO-coated slide was patterned with photoresist to feature a 500 μm diameter exposed region. The fluid cell (Figures 1B and 1C) was subsequently filled with a pH 12 solution with 1 mM luminol and 5 mM $H_2O_2$. When viewed using bright field microscopy, the region without photoresist was clearly visible as a circle (Figure 1D). Upon the application of a 1.6 V DC voltage across the fluid cell for 2 s with no external illumination, bright ECL was observed centered on the exposed ITO circle with effectively no ECL intensity present in the insulating region outside the circle. Interestingly, the border of the circle appeared slightly brighter, which suggested possible artifacts due to phenomena such as diffusion. Such artifacts suggests that while ECL may be easy to generate, converting this to a quantitative imaging approach may require further exploration.

While ECL was clearly visible in the conductive regions of the sample, it was not clear whether the chosen experimental conditions produced the largest ECL intensity or if they suffered from artifacts from processes such as reagent diffusion. In order to study whether the solution composition affected these considerations, we defined a method for quantifying ECL intensity. In particular, the intensity with no voltage applied was denoted a background intensity $\langle I_0 \rangle$ (Figure 2A). After the application of a voltage, the average intensity inside the circular region was denoted $\langle I_V \rangle$. Note that the outer 15% of the circular region was omitted from this analysis to avoid contributions from the bright ring at the edge of the feature (Figure 2B).

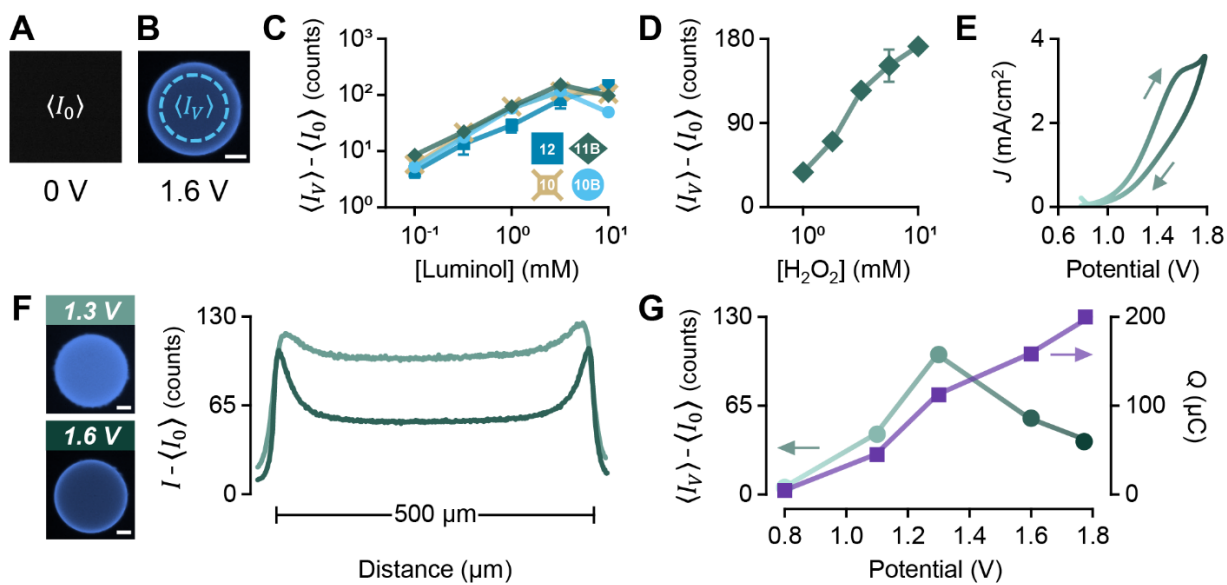

**Figure 2.** (A) Optical micrograph through the electrochemical cell with no external illumination and no voltage applied. This image is used to define an average dark intensity $\langle I_0 \rangle$. (B) Optical micrograph through the electrochemical cell with no external illumination but applied voltage $V = 1.6$ V. The bright circle corresponds to a 500 μm diameter region of exposed anode. The representative ECL intensity $\langle I_V \rangle$ is defined as the average intensity inside the dashed circle. (C) $\langle I_V \rangle - \langle I_0 \rangle$ vs. luminol concentration for four buffers with 5 mM hydrogen peroxide with $V = 1.6$ V applied for duration $t = 2$ s. Buffers studied are (12) pH 12 phosphate, (11B) pH 11 bicarbonate, (10) pH 10 borate, and (10B) pH 10 bicarbonate (full details in methods). (D) $\langle I_V \rangle - \langle I_0 \rangle$ vs. hydrogen peroxide concentration in bicarbonate buffer at pH 11 with 3.2 mM luminol with $V = 1.6$ V and $t = 2$ s. (E) Current density $J$ vs. $V$ taken using a two-terminal cyclic voltammogram of a bicarbonate buffer at pH 11 with 3.2 mM luminol and 5.6 mM hydrogen peroxide. (F) Micrograph intensity $I - \langle I_0 \rangle$ vs. position shown for two experiments taken with different $V$ with $t = 2$ s of a bicarbonate buffer at pH 11 with 3.2 mM luminol and 5.6 mM hydrogen peroxide. Full optical micrographs are shown for each condition on the left. (G) $\langle I_V \rangle - \langle I_0 \rangle$ and total charge passed $Q$ vs. $V$ taken simultaneously with $t = 2$ s for a bicarbonate buffer at pH 11 with 3.2 mM luminol and 5.6 mM hydrogen peroxide. All scale bars are 100 μm. All micrograph images were taken in greyscale and false-colored to represent the color seen by eye. Error bars for (C) and (D) represent standard deviation of three trials using different cells and are smaller than markers if not visible.

With a quantitative measure of ECL intensity in hand, we sought to explore the influence of processing conditions on ECL intensity. Prior work has studied a variety of factors including pH,[20–22] buffer composition,[21] luminol concentration,[20,21] and hydrogen peroxide concentration.[20] Thus, we tested the effects of each of these factors to determine the parameters that resulted in the highest ECL intensity. The buffer compositions and pHs tested included pH 10 sodium borate, pH 10 sodium bicarbonate, pH 11 sodium bicarbonate, and pH 12 disodium phosphate. Each of these buffer conditions were studied with luminol concentrations between 0.1 and 10 mM (Figure 2C). Three key features emerged from this exploration. (1) The choice of buffer is imperative with different buffers exhibiting more than a factor of 2 variation in ECL intensity with the same luminol concentration at the same buffer pH. (2) Below 3.2 mM luminol, all buffers exhibited an increase in ECL intensity with increasing luminol concentration, which suggests that, in this concentration range, the reaction is diffusion limited. (3) At higher luminol concentrations, the ECL intensity was dependent on buffer composition with some buffers leading to a decrease in ECL intensity at 10 mM luminol. These data are consistent with previous findings that show an optimal luminol concentration with higher concentrations leading to less ECL.[20,21] This decrease in ECL with increasing luminol concentration has been attributed to electrode passivation at high luminol concentrations.[21,26] Furthermore, it is well established that pH has a significant effect on ECL signal with alkaline pHs resulting in increasingly higher signals until a threshold value where ECL signal decreases.[20–22,24] Therefore, the solution conditions expected to result in the highest ECL intensity were pH 11 sodium bicarbonate buffer with 3.2 mM luminol.

Having identified optimized buffer conditions and luminol concentrations, identifying the optimal hydrogen peroxide concentration represented a balance between signal intensity and the robustness of the imaging method. In particular, it has been shown that high ECL intensity is observed with $H_2O_2$ in excess,[20,22–24] however, in our initial feasibility experiments, high concentrations of $H_2O_2$ led to bubble formation which obscured optical imaging and thus ECL signal. To optimize these competing considerations, ECL measurements were performed with peroxide concentrations between 1 and 10 mM using the optimized luminol solution (3.2 mM luminol in pH 11 bicarbonate buffer). While the ECL intensity did increase with increasing peroxide concentration (Figure 2D), only a 14% increase was observed in going from 5.6 to 10 mM. This suggests that these concentrations represented a suitable excess concentration relative to luminol such that the peroxide was not substantially limiting the reaction. Thus, 5.6 mM $H_2O_2$ was selected as a compromise to balance ECL intensity and bubble formation.

After establishing the optimal composition of the working solution, cyclic voltammetry (CV) was used to identify the functional voltage range for this solution composition and shed additional light on the electrochemical transformations taking place. Figure 2E shows a typical CV curve taken on a solution with 3.2 mM luminol, 5.6 mM $H_2O_2$, in pH 11 bicarbonate buffer. A shoulder is observed in the CV trace at ~1.4 V, suggesting mass-transfer limitation of the electrochemical reaction is taking place at this voltage and an additional oxidation at higher voltages, presumably the oxidation of hydrogen peroxide. To explore this reaction's relevance to ECL, we performed ECL imaging experiments on two samples, one collected with 1.3 V and the other with 1.6 V. While both exhibited bright circles of ECL, the center of the circular area was significantly dimmer in the sample at the higher voltage (shown in Figure 2F micrograph). To analyze the apparent ring in the image acquired at the higher voltage, the intensity profile $I$ was computed along a 550 µm long line through the center of the circle by averaging the pixel values in a 15 µm-wide region on either side of the line (Figure 2F). This analysis confirmed that not only did the 1.3 V image exhibit a higher average intensity in the circular region than the 1.6 V image, but that the ring artifact was markedly less intense at the lower voltage.

In order to more quantitatively study the connection between the electrical and the optical signal, ECL experiments were performed at a series of five voltages while observing the charge passed using chronoamperometry (Figure 2G). Comparing the intensity with charge passed $Q$, intensity increased with voltage with voltages ≤ 1.3 V and then decreased with increasing voltage for > 1.3 V. Since $Q$ monotonically increased with voltage, this indicates that other electrochemical processes were taking place that inhibited ECL. Given that this experiment is taking place in the presence of hydrogen peroxide, hydrogen peroxide oxidation is an obvious candidate reaction. Since the ring formation is more substantial in the presence of this competing reaction, it suggests that the depletion of hydrogen peroxide is inhibiting ECL more strongly in the center region where diffusion is limited.

Considering that the voltage-dependent experiments suggest that light intensity is non-monotonically dependent upon reagent concentration and is influenced by secondary reactions, it is interesting to consider whether exploring the dynamics of ECL could allow one to find a balance between imaging artifacts and signal intensity. To explore this, a series of experiments were performed in which the duration $t$ over which the voltage was applied was varied from 10 ms to 30 s (Figure 3A). Interestingly, the intensity measured at 1.3 V increased nearly linearly once the signal was larger than the noise intensity $I_n$, which is defined as the standard deviation of the dark image. Such a linear increase in signal with time is what one would expect if the reaction were proceeding light in a time-invariant fashion. In contrast, when measuring at 1.6 V, the signal drops below the linear power law at ~300 ms and adopts a $t^{1/2}$ power law, which is consistent with the reaction being limited by diffusion. These results suggest an explanation in which the 1.6 V process generates more ECL initially, but then shows a decrease in the rate of ECL generation due to hydrogen peroxide depletion at longer times. The spatial profiles (Figure 3B) of these experiments are consistent with this mechanism, in which a deviation from the linear trend is accompanied by the presence of the ring artifact. This ring can be explained as regions in which shorter diffusion is required to replenish the depleted hydrogen peroxide leading to higher ECL intensity.

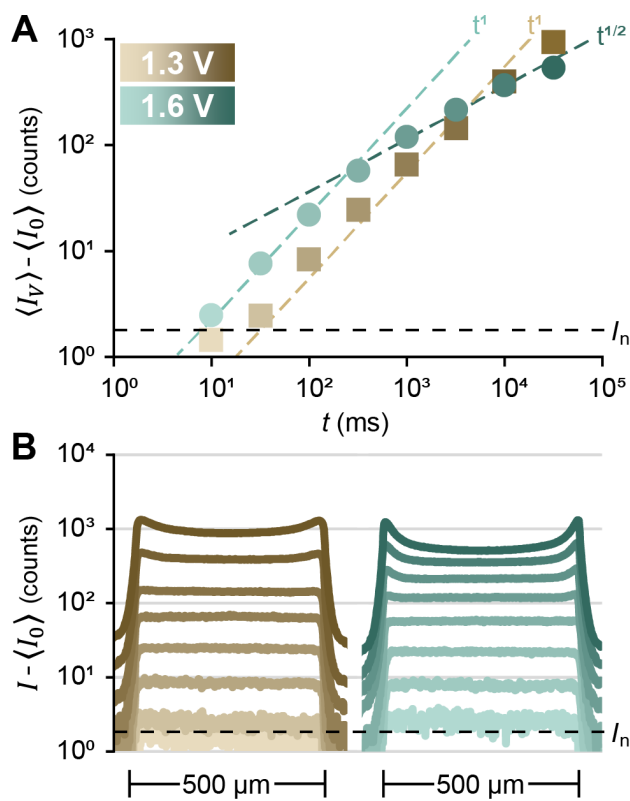

**Figure 3.** (A) $\langle I_V \rangle - \langle I_0 \rangle$ vs. $t$ for $V = 1.3$ V (brown) and $V = 1.6$ V (teal) of a bicarbonate buffer at pH 11 with 3.2 mM luminol and 5.6 mM hydrogen peroxide. The calculated noise ($I_n$) is indicated by the black dashed line. (B) Micrograph intensity $I - \langle I_0 \rangle$ vs. position shown for two experiments taken with different $V$ of a bicarbonate buffer at pH 11 with 3.2 mM luminol and 5.6 mM hydrogen peroxide. Color saturation increases with increasing time as shown in (A).

The results of the imaging and optimization experiments suggest a few important considerations for developing an ECL assay for defect detection. Critically, it is possible to choose conditions where local conductivity maps well to local ECL intensity. The best conditions for this are those which the traces in Figure 3B are highest while still being flat, or 3.2 s at 1.3 V. That said, if the goal is simply to generate the largest possible intensity so that small defects can readily be visualized, longer times can be used with 30 s at 1.3 V resulting in the highest absolute signal. Practically speaking, using as short a time as possible is still preferred in general as shorter times will use up less reagent, potentially allowing more images to be taken while reducing the time needed for each measurement.

While the prior experiments had focused on large conductive regions as a tool to optimize the assay conditions, a major goal of this work is to identify small defects in otherwise insulating films. To explore whether ECL could be used to identify sub-micron conductive defects in electrically insulating films, we performed a series of experiments in which an ITO-coated slide was coated with insulating PMMA and patterned using electron beam lithography (EBL). As the characteristic defects of interest are pinholes and cracks, we studied both lines that were designed to be between 0.1 to 1 µm wide and circles that were designed to be between 0.1 to 1 µm in diameter. The final dimensions of these lines and dots were measured using SEM and found to be between 0.44 – 1.46 µm wide and between 0.62 – 1.63 µm in diameter, respectively. ECL was performed on these samples using the optimized reagents, $V = 1.3$ V, and $t = 10$ s. The results of these ECL studies for both the line and dot features are shown in Figure 4A and Figure 4B as both ECL images and line cuts. All features were found to provide a clear optical signal with the shape of the ECL image matching the patterned shape.

While our lithography system was not readily capable of patterning substantially smaller features than those explored in Figures 4A and 4B, we hypothesized that these experiments could allow us to estimate the smallest features that would be possible to identify using this approach. In particular, we computed the average ECL intensity by integrating the intensity in a 10 µm square centered on each feature. Plotting this integrated intensity vs. line width (Figure 4C) revealed that there is a monotonic increase in intensity with increasing line width in a manner that is linear for lines narrower than 1 µm. Extrapolating to estimate the minimum resolvable line as the point where this fit line crosses the noise floor, we estimate that lines as narrow as 2.5 nm should be resolvable under these conditions. To estimate the minimum resolvable circular features, we integrated the intensity in a 5 µm circle centered on the dot of interest and found that this was highly linear with the dot area (Figure 4D), as one would expect. Extrapolating this curve to smaller areas, we find that circles with a radius of 35 nm should be the smallest resolvable features. For both the line and circle features, the background noise was estimated by calculating the integrated intensity in 15 different regions where no features were present in the ECL image and then subtracting the dark current (i.e., integrated intensity from when no voltage is applied). This led to lines having noise ~20 counts and circles having noise of 2.5 counts. Interestingly, when considering that the area of the sample interrogated by a single pixel of the camera is 1290 × 1290 $nm^2$ at 5× magnification, the area of the minimum resolvable line and circle features occupy very similar fractions of this area, with dots taking up 1/420 and lines taking up 1/516 of the area. This is consistent with the quantity of ECL generated being proportional to the exposed area.

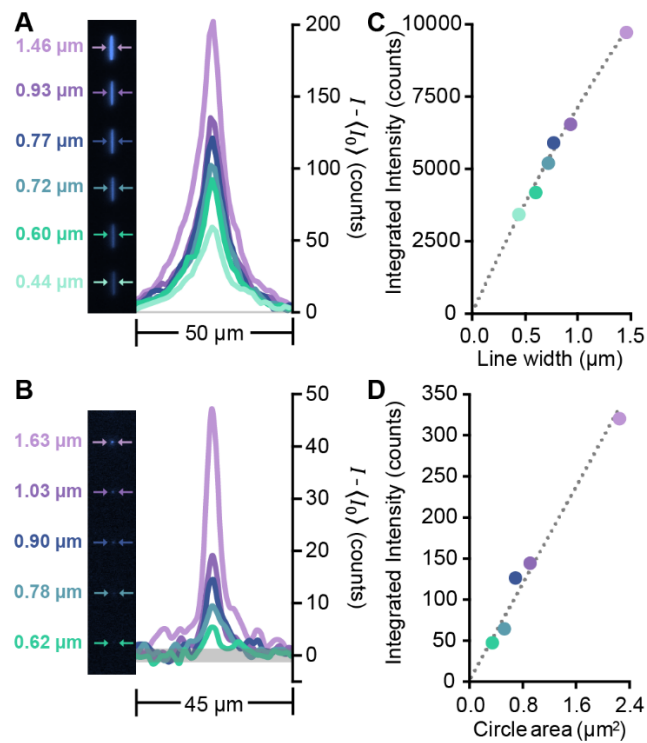

**Figure 4.** (A) ECL micrograph of a line array and line cuts of the micrograph intensity $I - \langle I_0 \rangle$ across each feature. ECL was carried out with $V = 1.3$ V for $t = 10$ s in a bicarbonate buffer at pH 11 with 3.2 mM luminol and 5.6 mM hydrogen peroxide. (B) ECL micrograph of circle array and line cuts of $I - \langle I_0 \rangle$ across each feature. ECL was carried out with $V = 1.3$ V for 10 s in a bicarbonate buffer at pH 11 with 3.2 mM luminol and 5.6 mM hydrogen peroxide. (C) Integrated intensity vs. line width calculated from the data in (A). (D) Integrated intensity vs circle area calculated from the data in (B). All micrograph images were taken in greyscale and false-colored to represent the color seen by eye. Grey bands in (A) and (B) represent the noise floor.

To explore the degree to which this assay could be used to provide insight about defects on nominally insulating films, we captured a series of ECL images on a 7.2 × 3.2 mm² region of an ITO-coated glass slide onto which an ultrathin layer of dielectric poly(phenylene oxide) had been electrodeposited.[27] These and similar ultrathin functional coatings have recently been shown to be of interest for advanced energy technologies.[3,28] As this region was much larger than the region corresponding to the field of view of the microscope, this imaging task was completed by iteratively taking ECL images and then moving the microscope stage. The motion was performed manually and there was a 10-20% areal overlap between neighboring frames to facilitate stitching the images together. Each ECL image was collected by applying 1.3 V for 10 s. The resulting images were stitched together to create a composite (Figure 5A).

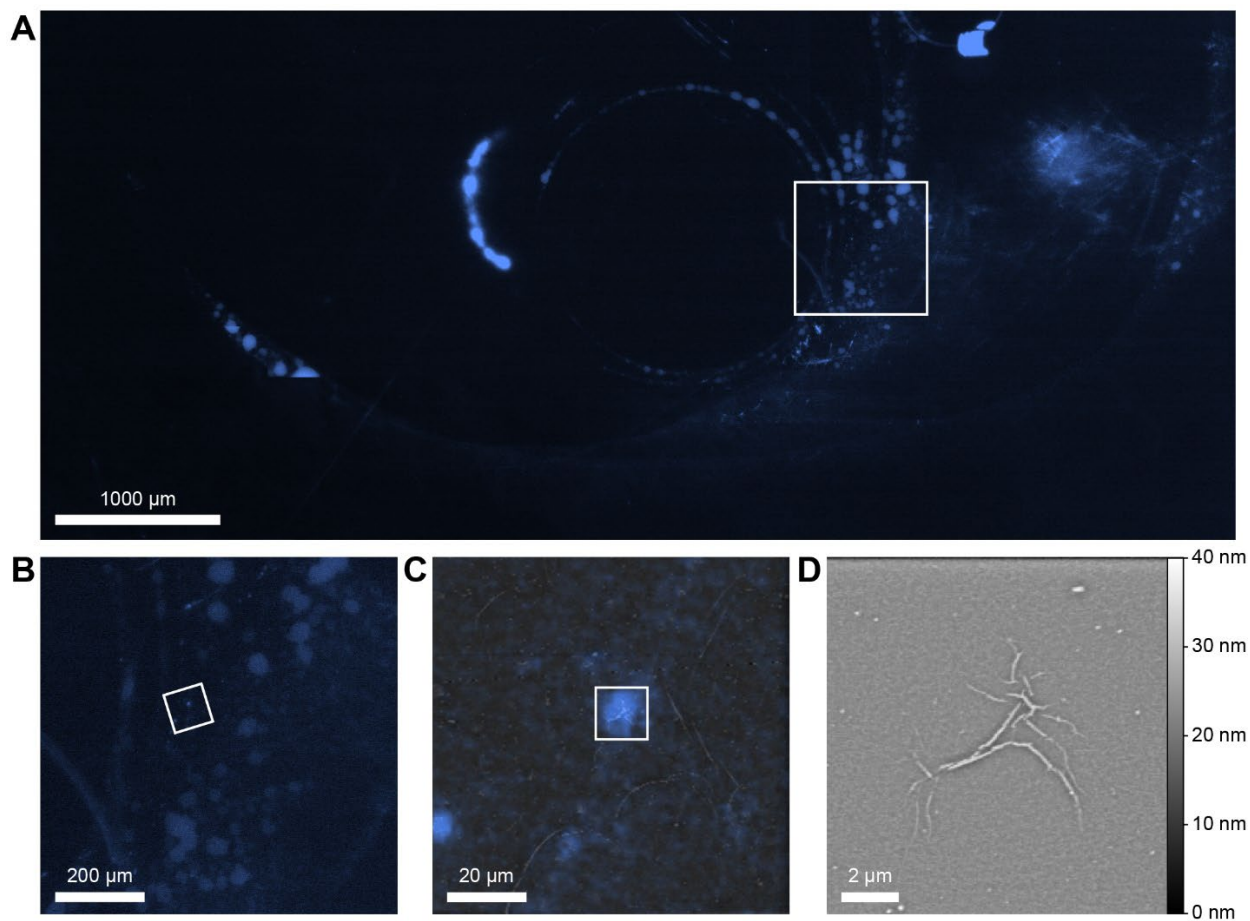

**Figure 5.** (A) Composite ECL micrograph of a phenol film taken with 5× objective, camera exposure 12 s, with 3.2 mM luminol, 5.6 mM $H_2O_2$, in pH 11 bicarbonate buffer with an applied potential of 1.3 V for 10 s for each frame. (B) Magnified view of the noted region in (A) showing a region of interest for further characterization. (C) Magnified and rotated view of the noted region in (B) overlaid on an AFM topographical image of the same location. (D) AFM topographical image corresponding to the noted region in (C). All ECL micrographs were taken in greyscale and false-colored to represent the color seen by eye.

Examining the large-area composite ECL of the phenol film, a few features were immediately apparent that provided important feedback for synthesis. First, it was clear that there were several large defects on the film including circular bands that were tens of microns wide. While these features were not visible in bright or dark-field optical microscopy, their circular shape leads us to believe that they were related to the process of a solvent drying on the sample surface. Analysis of the substrate preparation, film synthesis, and subsequent electrodeposition led us to conclude that the cleaning procedure used for substrate preparation was inadequate.

While the large-scale information that stemmed from the whole image is important and can provide a unique window into optimizing the deposition process, equally important is understanding the origin of individual defects. In particular, we sought to test whether bright spots in ECL could be linked to their nanoscale morphology to identify the structural origin of a defect. To explore this, we identified a region of interest (Figure 5B) with a particularly bright spot and co-localized this region using AFM through a combination of optically visible imperfections on the film surface as fiducial marks in addition to a grid system that was attached to the back of the sample. After locating the region of interest by taking several large format AFM images onto which the ECL images can be overlaid (Figure 5C), we performed a topographic image of the bright spot in the center of the region of interest. Interestingly, this $13 \times 13$ $\mu m^2$ AFM image revealed a cluster of sub-micron wrinkles in the film with prominent ridges between 100 – 250 nm in width (Figure 5D). This zoomed-in image suggests that this defect is not a large region that wasn't deposited or a piece of dirt that prevented deposition, but rather a result of film overgrowth or swelling leading to buckling on the film surface. This type of information can provide insight on defect formation and feedback for synthesis optimization.

The process of conducting a large-area scan on a nominally insulating film demonstrates the key utility of this ECL imaging method as a rapid approach for identifying small defects in large areas. Significantly, the ability to colocalize ECL features with AFM images provides a flexibility to efficiently study families of defects. Using the maximum field of view ($90 \times 90$ $\mu m^2$) of our AFM system to complete scans across the entire region imaged by ECL would have required scanning an area of 23,040,000 $\mu m^2$ or >2,800 scans. Not considering the burden of processing the data collected, the time needed to complete the experiment itself would exceed 900 hours. For comparison, the same area scanned using the optimized luminol-based ECL assay developed in this work required 18 ECL images and 18 reference images, with a camera exposure time of 12 s for each image and 30 s between voltage applications to allow equalization of reagents through diffusion, resulting in data collection within 20 minutes. Perhaps most importantly, even if the tedious AFM characterization was performed, it would not be obvious which defects were conductive while ECL imaging is directly measuring the functional property of interest.

**Conclusions**

Taken together, we have developed a novel method for detecting sub-micron defects in insulating thin films using low resolution tools. We performed a systematic study of reagent concentrations, voltages, and excitation time that allowed us to optimize reaction conditions. Interestingly, while luminol concentration, buffer composition, pH, and excitation voltage can be optimized to find a maximum signal, hydrogen peroxide concentration and excitation monotonically influence ECL signal but present tradeoffs in other ways. Specifically, excess hydrogen peroxide leads to bubble formation while prolonged measurement leads to diffusion-based artifacts. Taken together, this optimization directly improves the resolving power of this analytical method with lines down to 2.5 nm in width and pinholes as small as 35 nm in radius being detectable in principle. Further, we show multi-image stitching and co-registered mapping of nanoscale defects covering a >20 mm$^2$ sample. Collectively, these results show that this ECL imaging can be used to rapidly screen nominally insulating films for nanoscale defects, a critical capability for confident application of advanced materials in stretchable electronics, conformal coatings, and photovoltaic devices. Given the optical nature of this measurement, it can be readily combined with other non-destructive techniques for characterizing functional films. Further, this assay is inherently parallelizable and thus amenable to incorporation in an automated process for high-throughput screening of multifunctional films.

AUTHOR INFORMATION

**Corresponding Author**

* Correspondence should be addressed to Keith A. Brown (brownka@bu.edu).

**Author Contributions**

The manuscript was written through contributions of all authors. All authors have given approval to the final version of the manuscript. Project was conceived and guided by KAB and JGW, experiments and analysis were carried out by HQ, phenol films were synthesized by WW, figures were prepared by HQ.

**Funding Sources**

The authors acknowledge support from the National Science Foundation through CBET-2146597 and the Boston University College of Engineering Dean's Catalyst Award.

ACKNOWLEDGMENT

The authors thank the Boston University Photonics Center for providing access to instrumentation and resources critical to this work.